\documentclass[aip,reprint]{revtex4-1}
\usepackage[T1]{fontenc}
\usepackage{mathtools}
\usepackage{hyperref}
\usepackage{graphicx}
\usepackage{algorithm}
\usepackage{algorithmic}
\usepackage{verbatim}
\usepackage[euler]{textgreek}
\usepackage{xcolor}

\hypersetup{
	colorlinks,
	linkcolor={red!60!black},
	citecolor={red!60!black},
	urlcolor={red!80!black}
}

\newcommand{\vv}[1]{\mathbf{#1}}

\begin{document}

\title{AgentBasedModeling.jl: a tool for stochastic simulation of structured population dynamics}
\author{Paul Piho}
\author{Philipp Thomas}
\affiliation{Department of Mathematics, Imperial College London, London, UK.}

\begin{abstract}
\textbf{Summary:} 
Agent-based models capture heterogeneity among individuals in a population and are widely used
in studies of multi-cellular systems, disease, epidemics and demography to name a few. 
However, existing frameworks consider discrete time-step simulation or assume that agents' states only change as a result of discrete events.
In this note, we present \emph{AgentBasedModeling.jl}, a \emph{Julia} package for simulating stochastic agent-based population
models in continuous time. The tool allows to easily specify and simulate agents evolving through generic continuous-time jump-diffusions and interacting via continuous-rate processes. \emph{AgentBasedModeling.jl}
provides a powerful methodology for studying the effects of stochasticity on structured population dynamics. \\
\textbf{Availability:} \emph{AgentBasedModeling.jl} is a Julia package and available with the source code and usage
examples at \url{https://github.com/pihop/AgentBasedModeling.jl}.\\
\textbf{Contact:} \href{ppiho@imperial.ac.uk}{ppiho@imperial.ac.uk}\ \
\textbf{Corresponding author:} \href{p.thomas@imperial.ac.uk}{p.thomas@imperial.ac.uk}
\end{abstract}

\maketitle

\section{Introduction}
Interacting autonomous agents can exhibit complex dynamics requiring extensive computer simulation. Heterogeneity and
emergent phenomena in structured populations arise from actions and interactions of agents in response to their
characteristics. These unique characteristics manifest themselves as internal states that are subject to internal or
external fluctuations. Examples of agent-based systems include infected individuals in
epidemics~\citep{hoertel2020stochastica,didomenico2020impacta,kerr2021covasima,hinch2021openabmcovid19a}, cell
populations \citep{thomas2017making,garcia2018stochastic,ruess2023stochastic,piho2024feedbackb}, multi-cellular
systems~\citep{pleyer2023agent}, cancer cells in a growing
tumour~\citep{an2017optimizationa,cooper2020chastea,puccioni2024noiseinducedc},
corporations and government entities in economies~\citep{poledna2023economica,bertani2021complexitya,caiani2019doesa},
commuters in cities~\citep{nguyen2021overviewa} and people in social organisation dynamics~\citep{bruch2015agentbaseda}.
Simulation methods for these systems are gaining momentum as more and more data combines with computing power to
enable simulation-based inferences \citep{tankhilevich2019gpabc,cranmer2020frontier,jorgensen2022efficient,tang2023modelling}.

There exists a wealth of tools for the construction and simulation of agent-based models, such as
\emph{Agents.jl}~\citep{datseris2022agentsa}, MESA~\citep{kazil2020utilizinga},
Repast Simphony~\citep{north2013complexa} and NetLogo~\citep{wilensky1999netlogo}. These are based on incremental time
progression with a time step or probabilistically distributed event times. An implicit assumption in these approaches is
that agents' states do not change between events. A common discrete event simulation is provided by Gillespie's
algorithm~\citep{gillespie1976generala}, which applies only for unstructured models in which agents are indistinguishable. In the context of structured population dynamics that describe agents with different ages, life cycles, genetic differences, biochemical makeups or spatial
location~\citep{cushing1998introductiona}, approaches that capture both continuous time evolution of agent states and discrete events are needed.

Jump-diffusion processes~\citep{merton1976optiona} provide a generic framework for modelling agent state dynamics with single-agent algorithms implemented in standard packages such as
\emph{JumpProcesses.jl}~\citep{zagatti2024extendinga} or application-driven tools like
\emph{PyEcoLib}~\citep{nieto2023pyecoliba}. Structured populations of agents are either described by deterministic
partial differential equations~\citep{inaba2017agestructureda} or measure-valued stochastic
processes~\citep{donnelly1999particlea,bansaye2015stochastic}. In the agent-based world, such modelling considerations
have led to the development of tailored frameworks such as multi-scale models for cell
populations~\citep{matyjaszkiewicz2017bsima,dang2020cellulara,cooper2020chastea} where intracellular dynamics of cells evolve continuously specified by ordinary differential equations. However, general and extensible tools for structured populations, which couple stochastic state dynamics of interacting agents in continuous time, are currently limited.

\begin{figure*}[h!]
    \includegraphics[]{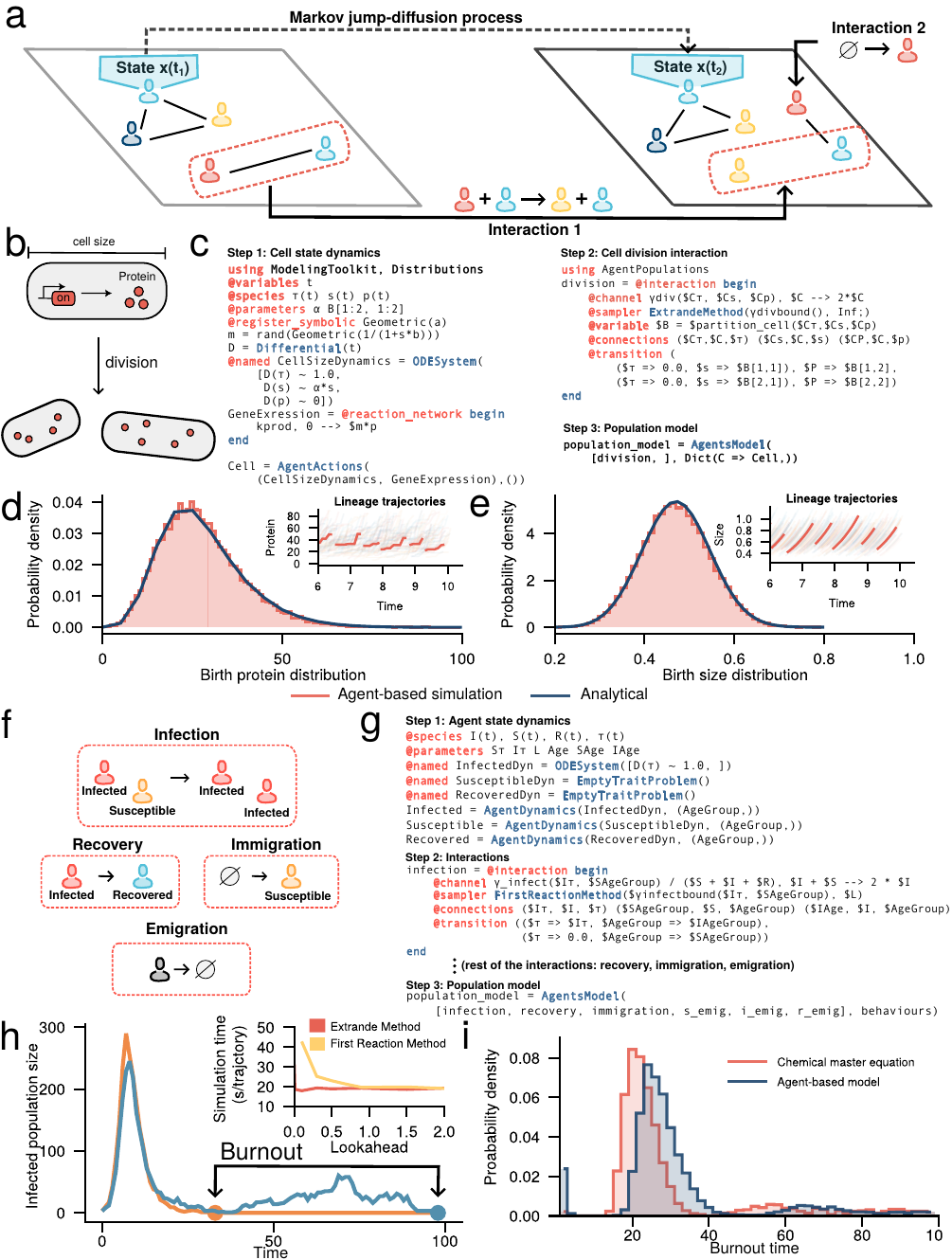}
    \caption[]{
        \textbf{(a)} Schematic representation of the agent-based models. The internal states of agents evolves according
        to a Markov jump-diffusion process. 
        \textbf{(b)} Cell division model. 
        \textbf{(c)} Model definition for the internal dynamics of a cell with size $s$ growing exponentially
            with growth rate $\alpha$ and stochastic bursty production of a protein $p$.         
        \textbf{(d-e)} Simulated birth protein and size lineage tree distributions of the cell division model (red)
        compared with the analytical computations (blue). Insets display the simulated lineage trajectories.
        \textbf{(f)} Agent interactions of a SIR model. 
        \textbf{(g)} Summary of model specification for the SIR.
        \textbf{(h)} Two sample trajectories of the agent-based model showing the epidemic burnout where the number
            of infected agents in a population becomes $0$.
            Inset shows the dependence of the simulation time for the agent-based model on the chosen lookup horizon.
        \textbf{(i}) Comparison of epidemic burnout distributions for the chemical master equation and agent-based
      }{}
    \label{fig:usg}
\end{figure*}

In this note, we introduce \emph{AgentBasedModeling.jl} allowing for easy specification and simulation of stochastic agent-based models where
internal agent dynamics are modelled as jump-diffusion processes and influence population-level interactions (illustrated in Fig~\ref{fig:usg}). 
We implement an exact simulation algorithm, allowing the rates of interactions between the agents to depend on the
continuously evolving internal agent states.
Our tool is integrated with the existing mathematical modelling libraries of
\emph{ModelingToolkit.jl}~\citep{ma2022modelingtoolkita} and \emph{Catalyst.jl}~\citep{loman2023catalystc} in \emph{Julia}
programming language for convenient specification of the internal state dynamical model of agents.

\section{Methods}
\subsection{Agent-based modelling}
We consider interactions between agents to depend on the agent type and their internal states, which evolve continuously in time. In a population of $M$ different agent types, $S_1, \ldots, S_M$, each agent is associated with a state vector $\vv{x}_{j}(t)$, $j \in 1, \ldots, M$ as a realisation of an autonomous Markov jump-diffusion process:
\begin{align}
	\mathrm{d} \vv{x}_j = \vv{f}_j(\vv{x}_j,t) \mathrm{d}t + B(\vv{x}_j(s))\mathrm{d}\vv{W}_j(t) + \sum_r \boldsymbol{\nu}_{rj}  \mathrm{d} {Y}_{rj},
	\label{eq:jumpdiff}
\end{align}
where $f_j$ is a deterministic drift vector corresponding to ordinary differential equation (ODE); $B(\vv{x}_j(t))d\vv{W}(t)$ is state-dependent Gaussian white noise corresponding to the stochastic differential equation part (SDE); and $d{Y}_{rj}$ are Poisson process jumps with path-dependent intensity $\int_0^t \mathrm{d}s w_{rj}(\vv{x}_j(s))$ and height $\boldsymbol{\nu}_{rj}$ that define the jump process.

We define an interaction rule that takes input agents of types $S_{n_1^c}, \ldots, S_{n_k^c}$ and creates output agents of
types $S_{m_1^c}, \ldots, S_{m_l^c}$, where $n_1^c, \ldots, n_k^c, m_1^c, \ldots, m_l^c \in \{1, \ldots, M\}$ are the indices of agent
types involved in the interaction. The interaction rule for all states $\vv{x}^-_{n_1^c},\ldots,\vv{x}^-_{n_k^c}$ is given by
\begin{align}
    & 
    S_{n_1^c}[\vv{x}^-_{n_1^c}] + \ldots + S_{n_k^c}[\vv{x}^-_{n_k^c}] 
    \nonumber \\ 
    & \hspace{4em} \xrightarrow{r_c\left(\vv{x}^-_{n_1^c}, \ldots, \vv{x}^-_{n_k^c}, t\right)}  S_{m_1^c}[\vv{x}^+_{m_1^c}] + \ldots + S_{m_l^c}[\vv{x}^+_{m_l^c}] 
    \nonumber \\
    & \vv{x}^+_{m_1^c}, \ldots, \vv{x}^+_{m_l^c} \sim B_c(\bullet|\vv{x}^-_{n_1^c},\ldots,\vv{x}^-_{n_l^c}). \label{eq:interaction}
\end{align}
The rates of the interactions $c$ depend on the internal states $\vv{x}^-_{n_1^c}, \ldots, \vv{x}^-_{n_k^c}$ of the
input agents and the states of the output agents are initialised probabilistically according to the transition kernel
$B_c$.
The transition kernel $B_c$ defines the probability of creating output agent states $\vv{x}^+_{m_1^c}, \ldots
\vv{x}^+_{m_l^c}$ given the input agent states $\vv{x}^-_{n_1^c}, \ldots \vv{x}^-_{n_k^c}$.

\subsection{Model specification}

\emph{AgentBasedModeling.jl} package allows for convenient specification and simulation of such models leveraging the
existing \emph{Julia} programming language ecosystem.
The first step (Fig~\ref{fig:usg}c~Step~1) in the specification is to define the internal state dynamics for all
agent types (Eq~\ref{eq:jumpdiff}). 
This is done as an ODE, SDE, jump process or combinations of it using
\emph{ModelingToolkit.jl}~\citep{ma2022modelingtoolkita} and \emph{Catalyst.jl}~\citep{loman2023catalystc}. 
The state dynamics are combined in the \verb!AgentDynamics! structure.

Interaction channels (Eq~\ref{eq:interaction}) are implemented with the \texttt{@interaction} macro environment
with the lines
\begin{align*}
    & \begin{aligned}
      \mathtt{@channel} \; r_c(\vv{x}^-_{n_1^c}, \ldots, \vv{x}^-_{n_k^c}, t), \; & S_{n_1^c} + \ldots + S_{n_k}^c \to \\  & S_{m_1^c} + \ldots + S_{m_l^c}
    \end{aligned} \\ 
    & \begin{aligned}
      \mathtt{@transition} \; (& {\vv{x}}_{m_1^c}(t) \Rightarrow \vv{x}^+_{m_1^c}, \ldots,
                               {\vv{x}}_{m_l^c}(t) \Rightarrow \vv{x}^+_{m_l^c}) \\
    \end{aligned} \\
    & \mathtt{@connection} \; (\vv{x}^-_{n_1^c}, S_{n_1^c}, \vv{x}_{n_1^c}(t)), \ldots, (\vv{x}^-_{n_k^c}, S_{n_k},
    \vv{x}_{n_k^c}(t))
\end{align*}
where ${\vv{x}}^+_{m_1^c}, \ldots, {\vv{x}}^+_{m_l^c}$ are sampled from the transition kernel
$B_c(\bullet | \vv{x}^-_{n_1^c}, \ldots, \vv{x}^-_{n_k^c})$.
The \texttt{@channel} line defines the rate function $r_c$ of the interaction and the interaction stoichiometry
while the $\texttt{@transition}$ line defines the initialisations of the output agents given the
transition kernel $B$. 
The \texttt{@connections} line is used to indicate which traits of which agents correspond to the symbols used in
the expressions.
The tuple $(\vv{x}^-_{n_1^c}, S_{n_1^c}, \vv{x}_{n_1^c}(t))$ denotes that the value of vector $\vv{x}^-_{n_1^c}$
corresponds to the state $\vv{x}_{n_1^c}(t)$ of the input agents of type $S_{n_1^c}$ at time $t$. These rules are matched to the individual instances of agents in the simulation.

As a step-by-step example, we consider a model of cell growth dynamics (illustration Fig~\ref{fig:usg}c) where exponential growth of cell size is coupled to bursty gene expression. Step 1 in Fig~\ref{fig:usg}c defines the internal dynamics of a cell via  \texttt{AgentActions}, where $\tau$ denotes time since the last division,
size $s$ growing exponentially with rate $\alpha$, and $p$ labels proteins expressed in geometrically distributed bursts. Step 2 defines the rate of the interaction channel via a user-defined function
\texttt{\textgamma div} taking the variables \texttt{C\texttau}, \texttt{Cs} and \texttt{Cp} that correspond to cell
age, size and protein counts respectively as inputs.
The \texttt{@transition} uses the helper variable $B$, computed by the
user-defined function \verb!partition_cell! that samples the transition kernel. Finally, in Step 3, the interactions and state dynamics are composed into a simulation model with the \verb!AgentsModel! function. 

\subsection{Simulation algorithm}
\emph{AgentBasedModeling.jl} provides a stochastic simulation algorithm to exactly simulate any agent-based model.
The outline of the algorithm is given as
Algorithm~\ref{alg:sim} and is based on the first reaction method for simulating Markov jump
processes~\citep{gillespie1977exacta}. 
To sample the next interaction time in a simulation time-interval $[t, t+\Delta t]$, the algorithm computes the
trajectories of agent states in that interval.
This is dependent on the model given for the agent state dynamics and can, for example, involve solving a system of ODEs, SDEs or sampling a trajectory of a jump-diffusion process for each agent in the population. 
We then construct a set of possible interaction instances between agents for each interaction channel. For each channel, the simulation algorithm samples and executes the instance with the fastest interaction time.

The package offers two algorithms for sampling the next interaction instance of a given channel, and different algorithms per channel can be combined within a simulation model. 
Simulation algorithms are specified within the \texttt{@interaction} macro using the \texttt{@sampler} keyword. The first method samples an interaction time for each instance of an interaction using the thinning
algorithm~\citep{lewis1979simulationb}.
This algorithm can be used by specifying the \verb!FirstInteractionMethod(bfn, L)! with a function \texttt{bfn} defining
the constant upper bound of the interaction rate and
lookahead horizon \texttt{L} defining how long the bound is valid for. The second algorithm uses the Extrande method~\citep{voliotis2016stochasticb} to sample the next interaction time for
each interaction channel and can be used by specifying \verb!ExtrandeMethod(bfn, L)! as the sampler with the same
user-defined functions as inputs. 

The simulation method \texttt{simulate} provides an option to save interaction times with the complete 
information about the corresponding interactions and the simulation state before the interaction.
The simulation state includes the state trajectories of the individual agents and thus collects all information needed
to provide a comprehensive picture of the simulated dynamics is available.

More fine-grained saving options are available for custom analysis of the results.
For example, the user can choose to save the trajectories of population snapshots, such as counts of agents of a given
type or distributions across the agents' states; and states of the input and output agents of interactions at the moment when the events took place.
As the tool is implemented as a package for the \emph{Julia} language we can utilise existing statistical and
visualisation packages for analyses.

\begin{algorithm}[H]
\caption{Stochastic simulator for the population models.}\label{alg:sim}
\begin{algorithmic}
    \REQUIRE Simulation time-span $[T_0, T]$, lookup horizon $\Delta t$.
    \STATE Let $t \leftarrow T_0$ and $t_w \leftarrow T_0 + \Delta t$ and consider the time window $[t, t_w]$.
    \STATE Let $\mathcal{S}$ be the initial population of agents at time $t$ defined by pairs of agent types and states
    at time $t$.
    \STATE For all agents in $\mathcal{S}$ simulate the state trajectories in the window $[t, t_w]$.
    \STATE Let $C$ be a set of interaction channels. For each $c \in C$ with input agent types $S_{n^{c}_1}, \cdots, S_{n^{c}_{k_c}}$
    construct the set $A_c$ of all without replacement combinations of agents in the population $\mathcal{S}$ that match
    the input types.
    \WHILE{$t \leq T$}
        \STATE For all channels $c$ let $f_c(a, s)$ be the interaction rate for a combination of agents $a$ at time
        $s \in [t, t_w]$ and let $\bar{f}_c(a)$ denote the upper bound of an interaction rate for all $s \in [t, t_w]$.
        Utilise one the following algorithms: the first reaction method where for each $a \in A_c$ use the thinning
        algorithm~\citep{lewis1979simulationb} with rate $f_c(a,s)$ and bound $\bar{f}_c(a)$ to sample next interaction
        times $t_a \in [t, t_w]$ and choose the time $t_c$ and input agents $a$ corresponding to $\mathsf{argmin}_{a}
        t_a$; 
        or apply the Extrande algorithm~\citep{voliotis2016stochasticb} with rate bound $\sum_r \bar{f}_c(a)$ to sample
        an interaction time $t_c$ and the input agents $a$. 
          \STATE Find the channel $c$ with the least next interaction time $\hat{t} = \mathsf{min}_c t_c$ and
          the corresponding combination of input agents $a_c$.
          \IF{$\hat{t} \leq t_w$}
              \STATE Given the set of input agents $a_c \subset \mathcal{S}$ with states $\vv{x}_{n_1^c}(\hat{t}), \ldots,
              \vv{x}_{n_{k_c}^c}(\hat{t})$ at time $\hat{t}$ and output agent types $S_{m_1^c}, \ldots, S_{m_{l_c}^c}$, 
              create output agents with the given types and states $\vv{x}^{+}_{m_1^c}, \ldots,
              \vv{x}^{+}_{m_{l_c}^c}$ at time $\hat{t}$ sampled from the kernel
              $B_c(\bullet|\vv{x}_{n_1^c}(\hat{t}), \ldots, \vv{x}_{n_{k_c}^c}(\hat{t}))$. Remove input agents $a_c$ from $\mathcal{S}$.
              \STATE Simulate the state trajectories of the output agents in the time interval $[\hat{t}, t_w]$, add the output agents to $\mathcal{S}$, and let $t \leftarrow \hat{t}$.
          \ELSE
              \STATE Let $t \leftarrow t_w $, $t_w \leftarrow t + \Delta t$ and simulate the state trajectories of all
              agents in $\mathcal{S}$ in time window $[t, t_w]$.
          \ENDIF
          \ENDWHILE
\end{algorithmic}
\end{algorithm}

\subsection{Applications to structured population dynamics}

We demonstrate the use of \emph{AgentBasedModeling.jl} with two models of varying complexity. We first show a cell
division model with a simple interaction structure and complex state models coupling intracellular reaction networks
with cell size and growth. Secondly we use a SIR model with simpler internal state dynamics but a rich network of interactions.

The cell division model is illustrated in Figure~\ref{fig:usg}b.
The cell state is defined by its age $\tau$, size $s$ and protein count $p$. The dynamics of protein count $p$ are given
by a Markov jump process modelling bursty protein production with the burst size depending on the cell size. Thus,
stochastic gene expression is coupled to cell size and growth. Figure~\ref{fig:usg}c demonstrates the specification of the dynamics of the agent state using a combination of 
\emph{ModelingToolkit.jl}, \emph{Catalyst.jl} and \emph{AgentBasedModeling.jl}.

Cell divisions are modelled via a single interaction channel
$C[\tau, s, p] \xrightarrow{r(\tau, s, p)} C[0, s_1, p_1] + C[0, s_2, p_2]$
with transition kernel $B(s_1, p_1, s_2, p_2 | s, p)$ defining the probability of the two daughter cells inheriting
sizes $s_1$, $s_2$ and protein counts $p_1$ and $p_2$ respectively, given the mother cell divides with size $s$ and
protein count $p$.
The functions used in the model to define division rate \texttt{\textgamma div} and partitioning \verb!partition_cell!
are user-defined \emph{Julia} functions. 

The stochastic simulation performed by \emph{AgentBasedModeling.jl} results in a lineage tree of cells.
Cell state dynamics corresponding to protein counts and cell size for the entire lineage tree are plotted in
Figure~\ref{fig:usg}d-e. We validated our implementation by computing birth protein distribution and size
analytically~\citep{thomas2018analysisa,thomas2021coordinationc}, which are in excellent agreement with our exact simulation algorithm.

The next application is a stochastic SIR model to study the influence of the incubation period  
on the probability of epidemic burnout~\citep{inaba2017agestructureda,parsons2024probabilitya}.
The model consists of three types of agents: susceptible (S), infected (I) and recovered (R).
The agent interactions are defined by infection, recovery, immigration and emigration interaction channels
(Fig~\ref{fig:usg}f). Immigration and emigration maintain the average population size at a steady state.  
Infected agents have a continuous internal state tracking the time since infection ($\tau$) that influences the rate of further infections and results in an incubation period. Each susceptible belongs to a fixed age group drawn from a distribution of age demographics (either older or younger) that further influences the infection rate. 

We obtained sample paths (Fig~\ref{fig:usg}h) from agent-based simulations showing burnout events where the disease is
stochastically eradicated. The burnout times have a multimodal distribution with peaks corresponding to eradication after multiple epidemic waves. Comparing these distributions to an unstructured Gillespie simulation (red), we find that the presence of an incubation period increases burnout probability initially but delays burnouts after the first and second outbreak waves (Fig~\ref{fig:usg}i). 

Simulating a large number of pairwise interactions is
time-consuming, and such computational burdens can only be partially alleviated through algorithmic choices
(Fig~\ref{fig:usg}h inset). These difficulties are common to all agent-based frameworks and could be diminished through
integration with coarse-graining methodologies such as in~\citep{nardini2021learninga}.

\section{Conclusion}

\emph{AgentBasedModeling.jl} provides a powerful tool for simulating structured population
dynamics in continuous time. Our approach models events via continuous rate functions coupled with agent state dynamics described by jump diffusions. Since both deterministic and stochastic internal state evolution are modelled, our tool applies to a range of agent-based applications, including single-cell or developed epidemic models. 

Existing agent-based simulation tools, such as \emph{Agents.jl}~\citep{datseris2022agentsa},
MESA~\citep{kazil2020utilizinga} and Repast Simphony~\citep{north2013complexa}, are often tailored to modelling spatial population structure. Our approach enables the simulation of structured population dynamics, which include spatial dynamics as a special case. Similarly, our tool extends Markov jump processes of single agents, such as provided by
\emph{PyEcoLib}~\citep{nieto2023pyecoliba} or \emph{JumpProcesses.jl}, to populations of interacting agents. Being implemented as a \emph{Julia} package, \emph{AgentBasedModeling.jl} allows computationally efficient modelling of agent-based populations without having to
implement a custom simulator for these frameworks, and it can make use of the existing frameworks for
parameter inference~\citep{tankhilevich2020gpabc} and data visualisation~\citep{danisch2021makiea}.

In summary, our tools make agent-based modelling and simulation of structured populations available to non-expert users
within a simple software package. Our approach will be useful for simulation-based inferences to advance our
understanding of the dynamics of interacting agents in biology, ecology, and social systems \citep{cranmer2020frontier}.


\section*{Funding}
UKRI supported this work through a Future Leaders Fellowship (MR/T018429/1 to PT).

\bibliography{text}

\begin{thebibliography}{46}%
\makeatletter
\providecommand \@ifxundefined [1]{%
 \@ifx{#1\undefined}
}%
\providecommand \@ifnum [1]{%
 \ifnum #1\expandafter \@firstoftwo
 \else \expandafter \@secondoftwo
 \fi
}%
\providecommand \@ifx [1]{%
 \ifx #1\expandafter \@firstoftwo
 \else \expandafter \@secondoftwo
 \fi
}%
\providecommand \natexlab [1]{#1}%
\providecommand \enquote  [1]{``#1''}%
\providecommand \bibnamefont  [1]{#1}%
\providecommand \bibfnamefont [1]{#1}%
\providecommand \citenamefont [1]{#1}%
\providecommand \href@noop [0]{\@secondoftwo}%
\providecommand \href [0]{\begingroup \@sanitize@url \@href}%
\providecommand \@href[1]{\@@startlink{#1}\@@href}%
\providecommand \@@href[1]{\endgroup#1\@@endlink}%
\providecommand \@sanitize@url [0]{\catcode `\\12\catcode `\$12\catcode
  `\&12\catcode `\#12\catcode `\^12\catcode `\_12\catcode `\%12\relax}%
\providecommand \@@startlink[1]{}%
\providecommand \@@endlink[0]{}%
\providecommand \url  [0]{\begingroup\@sanitize@url \@url }%
\providecommand \@url [1]{\endgroup\@href {#1}{\urlprefix }}%
\providecommand \urlprefix  [0]{URL }%
\providecommand \Eprint [0]{\href }%
\providecommand \doibase [0]{http://dx.doi.org/}%
\providecommand \selectlanguage [0]{\@gobble}%
\providecommand \bibinfo  [0]{\@secondoftwo}%
\providecommand \bibfield  [0]{\@secondoftwo}%
\providecommand \translation [1]{[#1]}%
\providecommand \BibitemOpen [0]{}%
\providecommand \bibitemStop [0]{}%
\providecommand \bibitemNoStop [0]{.\EOS\space}%
\providecommand \EOS [0]{\spacefactor3000\relax}%
\providecommand \BibitemShut  [1]{\csname bibitem#1\endcsname}%
\let\auto@bib@innerbib\@empty
\bibitem [{\citenamefont {Hoertel}\ \emph {et~al.}(2020)\citenamefont
  {Hoertel}, \citenamefont {Blachier}, \citenamefont {Blanco}, \citenamefont
  {Olfson}, \citenamefont {Massetti}, \citenamefont {Rico}, \citenamefont
  {Limosin},\ and\ \citenamefont {Leleu}}]{hoertel2020stochastica}%
  \BibitemOpen
  \bibfield  {author} {\bibinfo {author} {\bibfnamefont {N.}~\bibnamefont
  {Hoertel}}, \bibinfo {author} {\bibfnamefont {M.}~\bibnamefont {Blachier}},
  \bibinfo {author} {\bibfnamefont {C.}~\bibnamefont {Blanco}}, \bibinfo
  {author} {\bibfnamefont {M.}~\bibnamefont {Olfson}}, \bibinfo {author}
  {\bibfnamefont {M.}~\bibnamefont {Massetti}}, \bibinfo {author}
  {\bibfnamefont {M.~S.}\ \bibnamefont {Rico}}, \bibinfo {author}
  {\bibfnamefont {F.}~\bibnamefont {Limosin}}, \ and\ \bibinfo {author}
  {\bibfnamefont {H.}~\bibnamefont {Leleu}},\ }\bibfield  {title} {\enquote
  {\bibinfo {title} {A stochastic agent-based model of the {{SARS-CoV-2}}
  epidemic in {{France}}},}\ }\href {\doibase 10.1038/s41591-020-1001-6}
  {\bibfield  {journal} {\bibinfo  {journal} {Nat Med}\ }\textbf {\bibinfo
  {volume} {26}},\ \bibinfo {pages} {1417--1421} (\bibinfo {year}
  {2020})}\BibitemShut {NoStop}%
\bibitem [{\citenamefont {Di~Domenico}\ \emph {et~al.}(2020)\citenamefont
  {Di~Domenico}, \citenamefont {Pullano}, \citenamefont {Sabbatini},
  \citenamefont {Bo{\"e}lle},\ and\ \citenamefont
  {Colizza}}]{didomenico2020impacta}%
  \BibitemOpen
  \bibfield  {author} {\bibinfo {author} {\bibfnamefont {L.}~\bibnamefont
  {Di~Domenico}}, \bibinfo {author} {\bibfnamefont {G.}~\bibnamefont
  {Pullano}}, \bibinfo {author} {\bibfnamefont {C.~E.}\ \bibnamefont
  {Sabbatini}}, \bibinfo {author} {\bibfnamefont {P.-Y.}\ \bibnamefont
  {Bo{\"e}lle}}, \ and\ \bibinfo {author} {\bibfnamefont {V.}~\bibnamefont
  {Colizza}},\ }\bibfield  {title} {\enquote {\bibinfo {title} {Impact of
  lockdown on {{COVID-19}} epidemic in {{{\^I}le-de-France}} and possible exit
  strategies},}\ }\href {\doibase 10.1186/s12916-020-01698-4} {\bibfield
  {journal} {\bibinfo  {journal} {BMC Medicine}\ }\textbf {\bibinfo {volume}
  {18}},\ \bibinfo {pages} {240} (\bibinfo {year} {2020})}\BibitemShut
  {NoStop}%
\bibitem [{\citenamefont {Kerr}\ \emph {et~al.}(2021)\citenamefont {Kerr},
  \citenamefont {Stuart}, \citenamefont {Mistry}, \citenamefont {Abeysuriya},
  \citenamefont {Rosenfeld}, \citenamefont {Hart}, \citenamefont
  {N{\'u}{\~n}ez}, \citenamefont {Cohen}, \citenamefont {Selvaraj},
  \citenamefont {Hagedorn}, \citenamefont {George}, \citenamefont {Jastrz{\k
  e}bski}, \citenamefont {Izzo}, \citenamefont {Fowler}, \citenamefont
  {Palmer}, \citenamefont {Delport}, \citenamefont {Scott}, \citenamefont
  {Kelly}, \citenamefont {Bennette}, \citenamefont {Wagner}, \citenamefont
  {Chang}, \citenamefont {Oron}, \citenamefont {Wenger}, \citenamefont
  {{Panovska-Griffiths}}, \citenamefont {Famulare},\ and\ \citenamefont
  {Klein}}]{kerr2021covasima}%
  \BibitemOpen
  \bibfield  {author} {\bibinfo {author} {\bibfnamefont {C.~C.}\ \bibnamefont
  {Kerr}}, \bibinfo {author} {\bibfnamefont {R.~M.}\ \bibnamefont {Stuart}},
  \bibinfo {author} {\bibfnamefont {D.}~\bibnamefont {Mistry}}, \bibinfo
  {author} {\bibfnamefont {R.~G.}\ \bibnamefont {Abeysuriya}}, \bibinfo
  {author} {\bibfnamefont {K.}~\bibnamefont {Rosenfeld}}, \bibinfo {author}
  {\bibfnamefont {G.~R.}\ \bibnamefont {Hart}}, \bibinfo {author}
  {\bibfnamefont {R.~C.}\ \bibnamefont {N{\'u}{\~n}ez}}, \bibinfo {author}
  {\bibfnamefont {J.~A.}\ \bibnamefont {Cohen}}, \bibinfo {author}
  {\bibfnamefont {P.}~\bibnamefont {Selvaraj}}, \bibinfo {author}
  {\bibfnamefont {B.}~\bibnamefont {Hagedorn}}, \bibinfo {author}
  {\bibfnamefont {L.}~\bibnamefont {George}}, \bibinfo {author} {\bibfnamefont
  {M.}~\bibnamefont {Jastrz{\k e}bski}}, \bibinfo {author} {\bibfnamefont
  {A.~S.}\ \bibnamefont {Izzo}}, \bibinfo {author} {\bibfnamefont
  {G.}~\bibnamefont {Fowler}}, \bibinfo {author} {\bibfnamefont
  {A.}~\bibnamefont {Palmer}}, \bibinfo {author} {\bibfnamefont
  {D.}~\bibnamefont {Delport}}, \bibinfo {author} {\bibfnamefont
  {N.}~\bibnamefont {Scott}}, \bibinfo {author} {\bibfnamefont {S.~L.}\
  \bibnamefont {Kelly}}, \bibinfo {author} {\bibfnamefont {C.~S.}\ \bibnamefont
  {Bennette}}, \bibinfo {author} {\bibfnamefont {B.~G.}\ \bibnamefont
  {Wagner}}, \bibinfo {author} {\bibfnamefont {S.~T.}\ \bibnamefont {Chang}},
  \bibinfo {author} {\bibfnamefont {A.~P.}\ \bibnamefont {Oron}}, \bibinfo
  {author} {\bibfnamefont {E.~A.}\ \bibnamefont {Wenger}}, \bibinfo {author}
  {\bibfnamefont {J.}~\bibnamefont {{Panovska-Griffiths}}}, \bibinfo {author}
  {\bibfnamefont {M.}~\bibnamefont {Famulare}}, \ and\ \bibinfo {author}
  {\bibfnamefont {D.~J.}\ \bibnamefont {Klein}},\ }\bibfield  {title} {\enquote
  {\bibinfo {title} {Covasim: {{An}} agent-based model of {{COVID-19}} dynamics
  and interventions},}\ }\href {\doibase 10.1371/journal.pcbi.1009149}
  {\bibfield  {journal} {\bibinfo  {journal} {PLoS Comput. Biol.}\ }\textbf
  {\bibinfo {volume} {17}} (\bibinfo {year} {2021}),\
  10.1371/journal.pcbi.1009149}\BibitemShut {NoStop}%
\bibitem [{\citenamefont {Hinch}\ \emph {et~al.}(2021)\citenamefont {Hinch},
  \citenamefont {Probert}, \citenamefont {Nurtay}, \citenamefont {Kendall},
  \citenamefont {Wymant}, \citenamefont {Hall}, \citenamefont {Lythgoe},
  \citenamefont {Cruz}, \citenamefont {Zhao}, \citenamefont {Stewart},
  \citenamefont {Ferretti}, \citenamefont {Montero}, \citenamefont {Warren},
  \citenamefont {Mather}, \citenamefont {Abueg}, \citenamefont {Wu},
  \citenamefont {Legat}, \citenamefont {Bentley}, \citenamefont {Mead},
  \citenamefont {{Van-Vuuren}}, \citenamefont {{Feldner-Busztin}},
  \citenamefont {Ristori}, \citenamefont {Finkelstein}, \citenamefont
  {Bonsall}, \citenamefont {{Abeler-D{\"o}rner}},\ and\ \citenamefont
  {Fraser}}]{hinch2021openabmcovid19a}%
  \BibitemOpen
  \bibfield  {author} {\bibinfo {author} {\bibfnamefont {R.}~\bibnamefont
  {Hinch}}, \bibinfo {author} {\bibfnamefont {W.~J.~M.}\ \bibnamefont
  {Probert}}, \bibinfo {author} {\bibfnamefont {A.}~\bibnamefont {Nurtay}},
  \bibinfo {author} {\bibfnamefont {M.}~\bibnamefont {Kendall}}, \bibinfo
  {author} {\bibfnamefont {C.}~\bibnamefont {Wymant}}, \bibinfo {author}
  {\bibfnamefont {M.}~\bibnamefont {Hall}}, \bibinfo {author} {\bibfnamefont
  {K.}~\bibnamefont {Lythgoe}}, \bibinfo {author} {\bibfnamefont {A.~B.}\
  \bibnamefont {Cruz}}, \bibinfo {author} {\bibfnamefont {L.}~\bibnamefont
  {Zhao}}, \bibinfo {author} {\bibfnamefont {A.}~\bibnamefont {Stewart}},
  \bibinfo {author} {\bibfnamefont {L.}~\bibnamefont {Ferretti}}, \bibinfo
  {author} {\bibfnamefont {D.}~\bibnamefont {Montero}}, \bibinfo {author}
  {\bibfnamefont {J.}~\bibnamefont {Warren}}, \bibinfo {author} {\bibfnamefont
  {N.}~\bibnamefont {Mather}}, \bibinfo {author} {\bibfnamefont
  {M.}~\bibnamefont {Abueg}}, \bibinfo {author} {\bibfnamefont
  {N.}~\bibnamefont {Wu}}, \bibinfo {author} {\bibfnamefont {O.}~\bibnamefont
  {Legat}}, \bibinfo {author} {\bibfnamefont {K.}~\bibnamefont {Bentley}},
  \bibinfo {author} {\bibfnamefont {T.}~\bibnamefont {Mead}}, \bibinfo {author}
  {\bibfnamefont {K.}~\bibnamefont {{Van-Vuuren}}}, \bibinfo {author}
  {\bibfnamefont {D.}~\bibnamefont {{Feldner-Busztin}}}, \bibinfo {author}
  {\bibfnamefont {T.}~\bibnamefont {Ristori}}, \bibinfo {author} {\bibfnamefont
  {A.}~\bibnamefont {Finkelstein}}, \bibinfo {author} {\bibfnamefont {D.~G.}\
  \bibnamefont {Bonsall}}, \bibinfo {author} {\bibfnamefont {L.}~\bibnamefont
  {{Abeler-D{\"o}rner}}}, \ and\ \bibinfo {author} {\bibfnamefont
  {C.}~\bibnamefont {Fraser}},\ }\bibfield  {title} {\enquote {\bibinfo {title}
  {{{OpenABM-Covid19}}---{{An}} agent-based model for non-pharmaceutical
  interventions against {{COVID-19}} including contact tracing},}\ }\href
  {\doibase 10.1371/journal.pcbi.1009146} {\bibfield  {journal} {\bibinfo
  {journal} {PLOS Computational Biology}\ }\textbf {\bibinfo {volume} {17}},\
  \bibinfo {pages} {e1009146} (\bibinfo {year} {2021})}\BibitemShut {NoStop}%
\bibitem [{\citenamefont {Thomas}(2017)}]{thomas2017making}%
  \BibitemOpen
  \bibfield  {author} {\bibinfo {author} {\bibfnamefont {P.}~\bibnamefont
  {Thomas}},\ }\bibfield  {title} {\enquote {\bibinfo {title} {Making sense of
  snapshot data: Ergodic principle for clonal cell populations},}\ }\href@noop
  {} {\bibfield  {journal} {\bibinfo  {journal} {J. R. Soc. Interface}\
  }\textbf {\bibinfo {volume} {14}},\ \bibinfo {pages} {20170467} (\bibinfo
  {year} {2017})}\BibitemShut {NoStop}%
\bibitem [{\citenamefont {Garc{\'{\i}}a}\ \emph {et~al.}(2018)\citenamefont
  {Garc{\'{\i}}a}, \citenamefont {V{\'a}zquez}, \citenamefont {Teixeira},\ and\
  \citenamefont {Alonso}}]{garcia2018stochastic}%
  \BibitemOpen
  \bibfield  {author} {\bibinfo {author} {\bibfnamefont {M.~R.}\ \bibnamefont
  {Garc{\'{\i}}a}}, \bibinfo {author} {\bibfnamefont {J.~A.}\ \bibnamefont
  {V{\'a}zquez}}, \bibinfo {author} {\bibfnamefont {I.~G.}\ \bibnamefont
  {Teixeira}}, \ and\ \bibinfo {author} {\bibfnamefont {A.~A.}\ \bibnamefont
  {Alonso}},\ }\bibfield  {title} {\enquote {\bibinfo {title} {Stochastic
  individual-based modeling of bacterial growth and division using flow
  cytometry},}\ }\href@noop {} {\bibfield  {journal} {\bibinfo  {journal}
  {Front. Microbiol.}\ }\textbf {\bibinfo {volume} {8}},\ \bibinfo {pages}
  {2626} (\bibinfo {year} {2018})}\BibitemShut {NoStop}%
\bibitem [{\citenamefont {Ruess}, \citenamefont {Ballif},\ and\ \citenamefont
  {Aditya}(2023)}]{ruess2023stochastic}%
  \BibitemOpen
  \bibfield  {author} {\bibinfo {author} {\bibfnamefont {J.}~\bibnamefont
  {Ruess}}, \bibinfo {author} {\bibfnamefont {G.}~\bibnamefont {Ballif}}, \
  and\ \bibinfo {author} {\bibfnamefont {C.}~\bibnamefont {Aditya}},\
  }\bibfield  {title} {\enquote {\bibinfo {title} {Stochastic chemical kinetics
  of cell fate decision systems: {{From}} single cells to populations and
  back},}\ }\href@noop {} {\bibfield  {journal} {\bibinfo  {journal} {J. Chem.
  Phys.}\ }\textbf {\bibinfo {volume} {159}} (\bibinfo {year}
  {2023})}\BibitemShut {NoStop}%
\bibitem [{\citenamefont {Piho}\ and\ \citenamefont
  {Thomas}(2024)}]{piho2024feedbackb}%
  \BibitemOpen
  \bibfield  {author} {\bibinfo {author} {\bibfnamefont {P.}~\bibnamefont
  {Piho}}\ and\ \bibinfo {author} {\bibfnamefont {P.}~\bibnamefont {Thomas}},\
  }\bibfield  {title} {\enquote {\bibinfo {title} {Feedback between stochastic
  gene networks and population dynamics enables cellular decision-making},}\
  }\href {\doibase 10.1126/sciadv.adl4895} {\bibfield  {journal} {\bibinfo
  {journal} {Sci. Adv.}\ }\textbf {\bibinfo {volume} {10}},\ \bibinfo {pages}
  {eadl4895} (\bibinfo {year} {2024})}\BibitemShut {NoStop}%
\bibitem [{\citenamefont {Pleyer}\ and\ \citenamefont
  {Fleck}(2023)}]{pleyer2023agent}%
  \BibitemOpen
  \bibfield  {author} {\bibinfo {author} {\bibfnamefont {J.}~\bibnamefont
  {Pleyer}}\ and\ \bibinfo {author} {\bibfnamefont {C.}~\bibnamefont {Fleck}},\
  }\bibfield  {title} {\enquote {\bibinfo {title} {Agent-based models in
  cellular systems},}\ }\href@noop {} {\bibfield  {journal} {\bibinfo
  {journal} {Front. Phys.}\ }\textbf {\bibinfo {volume} {10}},\ \bibinfo
  {pages} {968409} (\bibinfo {year} {2023})}\BibitemShut {NoStop}%
\bibitem [{\citenamefont {An}\ \emph {et~al.}(2017)\citenamefont {An},
  \citenamefont {Fitzpatrick}, \citenamefont {Christley}, \citenamefont
  {Federico}, \citenamefont {Kanarek}, \citenamefont {Neilan}, \citenamefont
  {Oremland}, \citenamefont {Salinas}, \citenamefont {Laubenbacher},\ and\
  \citenamefont {Lenhart}}]{an2017optimizationa}%
  \BibitemOpen
  \bibfield  {author} {\bibinfo {author} {\bibfnamefont {G.}~\bibnamefont
  {An}}, \bibinfo {author} {\bibfnamefont {B.~G.}\ \bibnamefont {Fitzpatrick}},
  \bibinfo {author} {\bibfnamefont {S.}~\bibnamefont {Christley}}, \bibinfo
  {author} {\bibfnamefont {P.}~\bibnamefont {Federico}}, \bibinfo {author}
  {\bibfnamefont {A.}~\bibnamefont {Kanarek}}, \bibinfo {author} {\bibfnamefont
  {R.~M.}\ \bibnamefont {Neilan}}, \bibinfo {author} {\bibfnamefont
  {M.}~\bibnamefont {Oremland}}, \bibinfo {author} {\bibfnamefont
  {R.}~\bibnamefont {Salinas}}, \bibinfo {author} {\bibfnamefont
  {R.}~\bibnamefont {Laubenbacher}}, \ and\ \bibinfo {author} {\bibfnamefont
  {S.}~\bibnamefont {Lenhart}},\ }\bibfield  {title} {\enquote {\bibinfo
  {title} {Optimization and {{Control}} of {{Agent-Based Models}} in
  {{Biology}}: {{A Perspective}}},}\ }\href {\doibase
  10.1007/s11538-016-0225-6} {\bibfield  {journal} {\bibinfo  {journal} {Bull
  Math Biol}\ }\textbf {\bibinfo {volume} {79}},\ \bibinfo {pages} {63--87}
  (\bibinfo {year} {2017})}\BibitemShut {NoStop}%
\bibitem [{\citenamefont {Cooper}\ \emph {et~al.}(2020)\citenamefont {Cooper},
  \citenamefont {Baker}, \citenamefont {Bernabeu}, \citenamefont {Bordas},
  \citenamefont {Bowler}, \citenamefont {{Bueno-Orovio}}, \citenamefont
  {Byrne}, \citenamefont {Carapella}, \citenamefont {{Cardone-Noott}},
  \citenamefont {Cooper}, \citenamefont {Dutta}, \citenamefont {Evans},
  \citenamefont {Fletcher}, \citenamefont {Grogan}, \citenamefont {Guo},
  \citenamefont {Harvey}, \citenamefont {Hendrix}, \citenamefont {Kay},
  \citenamefont {Kursawe}, \citenamefont {Maini}, \citenamefont {McMillan},
  \citenamefont {Mirams}, \citenamefont {Osborne}, \citenamefont
  {Pathmanathan}, \citenamefont {{Pitt-Francis}}, \citenamefont {Robinson},
  \citenamefont {Rodriguez}, \citenamefont {Spiteri},\ and\ \citenamefont
  {Gavaghan}}]{cooper2020chastea}%
  \BibitemOpen
  \bibfield  {author} {\bibinfo {author} {\bibfnamefont {F.~R.}\ \bibnamefont
  {Cooper}}, \bibinfo {author} {\bibfnamefont {R.~E.}\ \bibnamefont {Baker}},
  \bibinfo {author} {\bibfnamefont {M.~O.}\ \bibnamefont {Bernabeu}}, \bibinfo
  {author} {\bibfnamefont {R.}~\bibnamefont {Bordas}}, \bibinfo {author}
  {\bibfnamefont {L.}~\bibnamefont {Bowler}}, \bibinfo {author} {\bibfnamefont
  {A.}~\bibnamefont {{Bueno-Orovio}}}, \bibinfo {author} {\bibfnamefont
  {H.~M.}\ \bibnamefont {Byrne}}, \bibinfo {author} {\bibfnamefont
  {V.}~\bibnamefont {Carapella}}, \bibinfo {author} {\bibfnamefont
  {L.}~\bibnamefont {{Cardone-Noott}}}, \bibinfo {author} {\bibfnamefont
  {J.}~\bibnamefont {Cooper}}, \bibinfo {author} {\bibfnamefont
  {S.}~\bibnamefont {Dutta}}, \bibinfo {author} {\bibfnamefont {B.~D.}\
  \bibnamefont {Evans}}, \bibinfo {author} {\bibfnamefont {A.~G.}\ \bibnamefont
  {Fletcher}}, \bibinfo {author} {\bibfnamefont {J.~A.}\ \bibnamefont
  {Grogan}}, \bibinfo {author} {\bibfnamefont {W.}~\bibnamefont {Guo}},
  \bibinfo {author} {\bibfnamefont {D.~G.}\ \bibnamefont {Harvey}}, \bibinfo
  {author} {\bibfnamefont {M.}~\bibnamefont {Hendrix}}, \bibinfo {author}
  {\bibfnamefont {D.}~\bibnamefont {Kay}}, \bibinfo {author} {\bibfnamefont
  {J.}~\bibnamefont {Kursawe}}, \bibinfo {author} {\bibfnamefont {P.~K.}\
  \bibnamefont {Maini}}, \bibinfo {author} {\bibfnamefont {B.}~\bibnamefont
  {McMillan}}, \bibinfo {author} {\bibfnamefont {G.~R.}\ \bibnamefont
  {Mirams}}, \bibinfo {author} {\bibfnamefont {J.~M.}\ \bibnamefont {Osborne}},
  \bibinfo {author} {\bibfnamefont {P.}~\bibnamefont {Pathmanathan}}, \bibinfo
  {author} {\bibfnamefont {J.~M.}\ \bibnamefont {{Pitt-Francis}}}, \bibinfo
  {author} {\bibfnamefont {M.}~\bibnamefont {Robinson}}, \bibinfo {author}
  {\bibfnamefont {B.}~\bibnamefont {Rodriguez}}, \bibinfo {author}
  {\bibfnamefont {R.~J.}\ \bibnamefont {Spiteri}}, \ and\ \bibinfo {author}
  {\bibfnamefont {D.~J.}\ \bibnamefont {Gavaghan}},\ }\bibfield  {title}
  {\enquote {\bibinfo {title} {Chaste: {{Cancer}}, {{Heart}} and {{Soft Tissue
  Environment}}},}\ }\href {\doibase 10.21105/joss.01848} {\bibfield  {journal}
  {\bibinfo  {journal} {J. Open Source Softw.}\ }\textbf {\bibinfo {volume}
  {5}},\ \bibinfo {pages} {1848} (\bibinfo {year} {2020})}\BibitemShut
  {NoStop}%
\bibitem [{\citenamefont {Puccioni}\ \emph {et~al.}(2024)\citenamefont
  {Puccioni}, \citenamefont {Pausch}, \citenamefont {Piho},\ and\ \citenamefont
  {Thomas}}]{puccioni2024noiseinducedc}%
  \BibitemOpen
  \bibfield  {author} {\bibinfo {author} {\bibfnamefont {F.}~\bibnamefont
  {Puccioni}}, \bibinfo {author} {\bibfnamefont {J.}~\bibnamefont {Pausch}},
  \bibinfo {author} {\bibfnamefont {P.}~\bibnamefont {Piho}}, \ and\ \bibinfo
  {author} {\bibfnamefont {P.}~\bibnamefont {Thomas}},\ }\href@noop {}
  {\enquote {\bibinfo {title} {Noise-induced survival resonances during
  fractional killing of cell populations},}\ } (\bibinfo {year} {2024}),\
  \Eprint {http://arxiv.org/abs/2402.19045} {arXiv:2402.19045 [physics, q-bio]}
  \BibitemShut {NoStop}%
\bibitem [{\citenamefont {Poledna}\ \emph {et~al.}(2023)\citenamefont
  {Poledna}, \citenamefont {Miess}, \citenamefont {Hommes},\ and\ \citenamefont
  {Rabitsch}}]{poledna2023economica}%
  \BibitemOpen
  \bibfield  {author} {\bibinfo {author} {\bibfnamefont {S.}~\bibnamefont
  {Poledna}}, \bibinfo {author} {\bibfnamefont {M.~G.}\ \bibnamefont {Miess}},
  \bibinfo {author} {\bibfnamefont {C.}~\bibnamefont {Hommes}}, \ and\ \bibinfo
  {author} {\bibfnamefont {K.}~\bibnamefont {Rabitsch}},\ }\bibfield  {title}
  {\enquote {\bibinfo {title} {Economic forecasting with an agent-based
  model},}\ }\href {\doibase 10.1016/j.euroecorev.2022.104306} {\bibfield
  {journal} {\bibinfo  {journal} {European Economic Review}\ }\textbf {\bibinfo
  {volume} {151}},\ \bibinfo {pages} {104306} (\bibinfo {year}
  {2023})}\BibitemShut {NoStop}%
\bibitem [{\citenamefont {Bertani}\ \emph {et~al.}(2021)\citenamefont
  {Bertani}, \citenamefont {Ponta}, \citenamefont {Raberto}, \citenamefont
  {Teglio},\ and\ \citenamefont {Cincotti}}]{bertani2021complexitya}%
  \BibitemOpen
  \bibfield  {author} {\bibinfo {author} {\bibfnamefont {F.}~\bibnamefont
  {Bertani}}, \bibinfo {author} {\bibfnamefont {L.}~\bibnamefont {Ponta}},
  \bibinfo {author} {\bibfnamefont {M.}~\bibnamefont {Raberto}}, \bibinfo
  {author} {\bibfnamefont {A.}~\bibnamefont {Teglio}}, \ and\ \bibinfo {author}
  {\bibfnamefont {S.}~\bibnamefont {Cincotti}},\ }\bibfield  {title} {\enquote
  {\bibinfo {title} {The complexity of the intangible digital economy: An
  agent-based model},}\ }\href {\doibase 10.1016/j.jbusres.2020.03.041}
  {\bibfield  {journal} {\bibinfo  {journal} {Journal of Business Research}\
  }\textbf {\bibinfo {volume} {129}},\ \bibinfo {pages} {527--540} (\bibinfo
  {year} {2021})}\BibitemShut {NoStop}%
\bibitem [{\citenamefont {Caiani}, \citenamefont {Russo},\ and\ \citenamefont
  {Gallegati}(2019)}]{caiani2019doesa}%
  \BibitemOpen
  \bibfield  {author} {\bibinfo {author} {\bibfnamefont {A.}~\bibnamefont
  {Caiani}}, \bibinfo {author} {\bibfnamefont {A.}~\bibnamefont {Russo}}, \
  and\ \bibinfo {author} {\bibfnamefont {M.}~\bibnamefont {Gallegati}},\
  }\bibfield  {title} {\enquote {\bibinfo {title} {Does inequality hamper
  innovation and growth? {{An AB-SFC}} analysis},}\ }\href {\doibase
  10.1007/s00191-018-0554-8} {\bibfield  {journal} {\bibinfo  {journal} {J Evol
  Econ}\ }\textbf {\bibinfo {volume} {29}},\ \bibinfo {pages} {177--228}
  (\bibinfo {year} {2019})}\BibitemShut {NoStop}%
\bibitem [{\citenamefont {Nguyen}\ \emph {et~al.}(2021)\citenamefont {Nguyen},
  \citenamefont {Powers}, \citenamefont {Urquhart}, \citenamefont
  {Farrenkopf},\ and\ \citenamefont {Guckert}}]{nguyen2021overviewa}%
  \BibitemOpen
  \bibfield  {author} {\bibinfo {author} {\bibfnamefont {J.}~\bibnamefont
  {Nguyen}}, \bibinfo {author} {\bibfnamefont {S.~T.}\ \bibnamefont {Powers}},
  \bibinfo {author} {\bibfnamefont {N.}~\bibnamefont {Urquhart}}, \bibinfo
  {author} {\bibfnamefont {T.}~\bibnamefont {Farrenkopf}}, \ and\ \bibinfo
  {author} {\bibfnamefont {M.}~\bibnamefont {Guckert}},\ }\bibfield  {title}
  {\enquote {\bibinfo {title} {An overview of agent-based traffic
  simulators},}\ }\href {\doibase 10.1016/j.trip.2021.100486} {\bibfield
  {journal} {\bibinfo  {journal} {Transportation Research Interdisciplinary
  Perspectives}\ }\textbf {\bibinfo {volume} {12}},\ \bibinfo {pages} {100486}
  (\bibinfo {year} {2021})}\BibitemShut {NoStop}%
\bibitem [{\citenamefont {Bruch}\ and\ \citenamefont
  {Atwell}(2015)}]{bruch2015agentbaseda}%
  \BibitemOpen
  \bibfield  {author} {\bibinfo {author} {\bibfnamefont {E.}~\bibnamefont
  {Bruch}}\ and\ \bibinfo {author} {\bibfnamefont {J.}~\bibnamefont {Atwell}},\
  }\bibfield  {title} {\enquote {\bibinfo {title} {Agent-{{Based Models}} in
  {{Empirical Social Research}}},}\ }\href {\doibase 10.1177/0049124113506405}
  {\bibfield  {journal} {\bibinfo  {journal} {Sociological Methods \&
  Research}\ }\textbf {\bibinfo {volume} {44}},\ \bibinfo {pages} {186--221}
  (\bibinfo {year} {2015})}\BibitemShut {NoStop}%
\bibitem [{\citenamefont {Tankhilevich}\ \emph {et~al.}(2019)\citenamefont
  {Tankhilevich}, \citenamefont {{Ish-Horowicz}}, \citenamefont {Hameed},
  \citenamefont {Roesch}, \citenamefont {Kleijn}, \citenamefont {Stumpf},\ and\
  \citenamefont {He}}]{tankhilevich2019gpabc}%
  \BibitemOpen
  \bibfield  {author} {\bibinfo {author} {\bibfnamefont {E.}~\bibnamefont
  {Tankhilevich}}, \bibinfo {author} {\bibfnamefont {J.}~\bibnamefont
  {{Ish-Horowicz}}}, \bibinfo {author} {\bibfnamefont {T.}~\bibnamefont
  {Hameed}}, \bibinfo {author} {\bibfnamefont {E.}~\bibnamefont {Roesch}},
  \bibinfo {author} {\bibfnamefont {I.}~\bibnamefont {Kleijn}}, \bibinfo
  {author} {\bibfnamefont {M.~P.}\ \bibnamefont {Stumpf}}, \ and\ \bibinfo
  {author} {\bibfnamefont {F.}~\bibnamefont {He}},\ }\href {\doibase
  10.1101/769299} {\enquote {\bibinfo {title} {{{GpABC}}: A {{Julia}} package
  for approximate {{Bayesian}} computation with {{Gaussian}} process
  emulation},}\ } (\bibinfo {year} {2019})\BibitemShut {NoStop}%
\bibitem [{\citenamefont {Cranmer}, \citenamefont {Brehmer},\ and\
  \citenamefont {Louppe}(2020)}]{cranmer2020frontier}%
  \BibitemOpen
  \bibfield  {author} {\bibinfo {author} {\bibfnamefont {K.}~\bibnamefont
  {Cranmer}}, \bibinfo {author} {\bibfnamefont {J.}~\bibnamefont {Brehmer}}, \
  and\ \bibinfo {author} {\bibfnamefont {G.}~\bibnamefont {Louppe}},\
  }\bibfield  {title} {\enquote {\bibinfo {title} {The frontier of
  simulation-based inference},}\ }\href@noop {} {\bibfield  {journal} {\bibinfo
   {journal} {Proc. Natl. Acad. Sci.}\ }\textbf {\bibinfo {volume} {117}},\
  \bibinfo {pages} {30055--30062} (\bibinfo {year} {2020})}\BibitemShut
  {NoStop}%
\bibitem [{\citenamefont {J{\o}rgensen}\ \emph {et~al.}(2022)\citenamefont
  {J{\o}rgensen}, \citenamefont {Ghosh}, \citenamefont {Sturrock},\ and\
  \citenamefont {Shahrezaei}}]{jorgensen2022efficient}%
  \BibitemOpen
  \bibfield  {author} {\bibinfo {author} {\bibfnamefont {A.~C.~S.}\
  \bibnamefont {J{\o}rgensen}}, \bibinfo {author} {\bibfnamefont
  {A.}~\bibnamefont {Ghosh}}, \bibinfo {author} {\bibfnamefont
  {M.}~\bibnamefont {Sturrock}}, \ and\ \bibinfo {author} {\bibfnamefont
  {V.}~\bibnamefont {Shahrezaei}},\ }\bibfield  {title} {\enquote {\bibinfo
  {title} {Efficient {{Bayesian}} inference for stochastic agent-based
  models},}\ }\href@noop {} {\bibfield  {journal} {\bibinfo  {journal} {PLoS
  Comput. Biol.}\ }\textbf {\bibinfo {volume} {18}},\ \bibinfo {pages}
  {e1009508} (\bibinfo {year} {2022})}\BibitemShut {NoStop}%
\bibitem [{\citenamefont {Tang}\ \emph {et~al.}(2023)\citenamefont {Tang},
  \citenamefont {J{\o}rgensen}, \citenamefont {Marguerat}, \citenamefont
  {Thomas},\ and\ \citenamefont {Shahrezaei}}]{tang2023modelling}%
  \BibitemOpen
  \bibfield  {author} {\bibinfo {author} {\bibfnamefont {W.}~\bibnamefont
  {Tang}}, \bibinfo {author} {\bibfnamefont {A.~C.~S.}\ \bibnamefont
  {J{\o}rgensen}}, \bibinfo {author} {\bibfnamefont {S.}~\bibnamefont
  {Marguerat}}, \bibinfo {author} {\bibfnamefont {P.}~\bibnamefont {Thomas}}, \
  and\ \bibinfo {author} {\bibfnamefont {V.}~\bibnamefont {Shahrezaei}},\
  }\bibfield  {title} {\enquote {\bibinfo {title} {Modelling capture efficiency
  of single-cell {{RNA-sequencing}} data improves inference of
  transcriptome-wide burst kinetics},}\ }\href@noop {} {\bibfield  {journal}
  {\bibinfo  {journal} {Bioinformatics}\ }\textbf {\bibinfo {volume} {39}},\
  \bibinfo {pages} {btad395} (\bibinfo {year} {2023})}\BibitemShut {NoStop}%
\bibitem [{\citenamefont {Datseris}, \citenamefont {Vahdati},\ and\
  \citenamefont {DuBois}(2022)}]{datseris2022agentsa}%
  \BibitemOpen
  \bibfield  {author} {\bibinfo {author} {\bibfnamefont {G.}~\bibnamefont
  {Datseris}}, \bibinfo {author} {\bibfnamefont {A.~R.}\ \bibnamefont
  {Vahdati}}, \ and\ \bibinfo {author} {\bibfnamefont {T.~C.}\ \bibnamefont
  {DuBois}},\ }\bibfield  {title} {\enquote {\bibinfo {title} {Agents.jl: A
  performant and feature-full agent-based modeling software of minimal code
  complexity},}\ }\href {\doibase 10.1177/00375497211068820} {\bibfield
  {journal} {\bibinfo  {journal} {SIMULATION}\ ,\ \bibinfo {pages}
  {00375497211068820}} (\bibinfo {year} {2022})}\BibitemShut {NoStop}%
\bibitem [{\citenamefont {Kazil}, \citenamefont {Masad},\ and\ \citenamefont
  {Crooks}(2020)}]{kazil2020utilizinga}%
  \BibitemOpen
  \bibfield  {author} {\bibinfo {author} {\bibfnamefont {J.}~\bibnamefont
  {Kazil}}, \bibinfo {author} {\bibfnamefont {D.}~\bibnamefont {Masad}}, \ and\
  \bibinfo {author} {\bibfnamefont {A.}~\bibnamefont {Crooks}},\ }\bibfield
  {title} {\enquote {\bibinfo {title} {Utilizing {{Python}} for {{Agent-Based
  Modeling}}: {{The Mesa Framework}}},}\ }in\ \href {\doibase
  10.1007/978-3-030-61255-9_30} {\emph {\bibinfo {booktitle} {Soc. {{Cult}}.
  {{Behav}}. {{Model}}.}}}\ (\bibinfo  {publisher} {Springer International
  Publishing},\ \bibinfo {address} {Cham},\ \bibinfo {year} {2020})\ pp.\
  \bibinfo {pages} {308--317}\BibitemShut {NoStop}%
\bibitem [{\citenamefont {North}\ \emph {et~al.}(2013)\citenamefont {North},
  \citenamefont {Collier}, \citenamefont {Ozik}, \citenamefont {Tatara},
  \citenamefont {Macal}, \citenamefont {Bragen},\ and\ \citenamefont
  {Sydelko}}]{north2013complexa}%
  \BibitemOpen
  \bibfield  {author} {\bibinfo {author} {\bibfnamefont {M.~J.}\ \bibnamefont
  {North}}, \bibinfo {author} {\bibfnamefont {N.~T.}\ \bibnamefont {Collier}},
  \bibinfo {author} {\bibfnamefont {J.}~\bibnamefont {Ozik}}, \bibinfo {author}
  {\bibfnamefont {E.~R.}\ \bibnamefont {Tatara}}, \bibinfo {author}
  {\bibfnamefont {C.~M.}\ \bibnamefont {Macal}}, \bibinfo {author}
  {\bibfnamefont {M.}~\bibnamefont {Bragen}}, \ and\ \bibinfo {author}
  {\bibfnamefont {P.}~\bibnamefont {Sydelko}},\ }\bibfield  {title} {\enquote
  {\bibinfo {title} {Complex adaptive systems modeling with {{Repast
  Simphony}}},}\ }\href {\doibase 10.1186/2194-3206-1-3} {\bibfield  {journal}
  {\bibinfo  {journal} {Complex Adapt Syst Model}\ }\textbf {\bibinfo {volume}
  {1}},\ \bibinfo {pages} {3} (\bibinfo {year} {2013})}\BibitemShut {NoStop}%
\bibitem [{\citenamefont {Wilensky}(1999)}]{wilensky1999netlogo}%
  \BibitemOpen
  \bibfield  {author} {\bibinfo {author} {\bibfnamefont {U.}~\bibnamefont
  {Wilensky}},\ }\href@noop {} {\enquote {\bibinfo {title} {{{NetLogo}}},}\
  }\bibinfo {howpublished} {Center for Connected Learning and Computer-Based
  Modeling, Northwestern University. Evanston, IL} (\bibinfo {year}
  {1999})\BibitemShut {NoStop}%
\bibitem [{\citenamefont {Gillespie}(1976)}]{gillespie1976generala}%
  \BibitemOpen
  \bibfield  {author} {\bibinfo {author} {\bibfnamefont {D.~T.}\ \bibnamefont
  {Gillespie}},\ }\bibfield  {title} {\enquote {\bibinfo {title} {A general
  method for numerically simulating the stochastic time evolution of coupled
  chemical reactions},}\ }\href {\doibase 10.1016/0021-9991(76)90041-3}
  {\bibfield  {journal} {\bibinfo  {journal} {Journal of Computational
  Physics}\ }\textbf {\bibinfo {volume} {22}},\ \bibinfo {pages} {403--434}
  (\bibinfo {year} {1976})}\BibitemShut {NoStop}%
\bibitem [{\citenamefont {Cushing}(1998)}]{cushing1998introductiona}%
  \BibitemOpen
  \bibfield  {author} {\bibinfo {author} {\bibfnamefont {J.~M.}\ \bibnamefont
  {Cushing}},\ }\href@noop {} {\emph {\bibinfo {title} {An Introduction to
  Structured Population Dynamics: Outgrowth of a Series of Lectures given at a
  Conference Held at {{North Carolina University}}, {{Raleigh}}, during
  {{June}} of 1997}}},\ \bibinfo {series} {{{CBMS-NSF}} Regional Conference
  Series in Applied Mathematics}\ No.~\bibinfo {number} {71}\ (\bibinfo
  {publisher} {{Soc. Industrial and Applied Mathematics}},\ \bibinfo {address}
  {Philadelphia, Pa},\ \bibinfo {year} {1998})\BibitemShut {NoStop}%
\bibitem [{\citenamefont {Merton}(1976)}]{merton1976optiona}%
  \BibitemOpen
  \bibfield  {author} {\bibinfo {author} {\bibfnamefont {R.~C.}\ \bibnamefont
  {Merton}},\ }\bibfield  {title} {\enquote {\bibinfo {title} {Option pricing
  when underlying stock returns are discontinuous},}\ }\href {\doibase
  10.1016/0304-405X(76)90022-2} {\bibfield  {journal} {\bibinfo  {journal}
  {Journal of Financial Economics}\ }\textbf {\bibinfo {volume} {3}},\ \bibinfo
  {pages} {125--144} (\bibinfo {year} {1976})}\BibitemShut {NoStop}%
\bibitem [{\citenamefont {Zagatti}\ \emph {et~al.}(2024)\citenamefont
  {Zagatti}, \citenamefont {Isaacson}, \citenamefont {Rackauckas},
  \citenamefont {Ilin}, \citenamefont {Ng},\ and\ \citenamefont
  {Bressan}}]{zagatti2024extendinga}%
  \BibitemOpen
  \bibfield  {author} {\bibinfo {author} {\bibfnamefont {G.~A.}\ \bibnamefont
  {Zagatti}}, \bibinfo {author} {\bibfnamefont {S.~A.}\ \bibnamefont
  {Isaacson}}, \bibinfo {author} {\bibfnamefont {C.}~\bibnamefont
  {Rackauckas}}, \bibinfo {author} {\bibfnamefont {V.}~\bibnamefont {Ilin}},
  \bibinfo {author} {\bibfnamefont {S.-K.}\ \bibnamefont {Ng}}, \ and\ \bibinfo
  {author} {\bibfnamefont {S.}~\bibnamefont {Bressan}},\ }\bibfield  {title}
  {\enquote {\bibinfo {title} {Extending {{JumpProcesses}}.jl for fast point
  process simulation with time-varying intensities},}\ }\href {\doibase
  10.21105/jcon.00133} {\bibfield  {journal} {\bibinfo  {journal} {Proc.
  JuliaCon Conf.}\ }\textbf {\bibinfo {volume} {6}},\ \bibinfo {pages} {133}
  (\bibinfo {year} {2024})}\BibitemShut {NoStop}%
\bibitem [{\citenamefont {Nieto}\ \emph {et~al.}(2023)\citenamefont {Nieto},
  \citenamefont {Blanco}, \citenamefont {{Vargas-Garc{\'i}a}}, \citenamefont
  {Singh},\ and\ \citenamefont {Manuel}}]{nieto2023pyecoliba}%
  \BibitemOpen
  \bibfield  {author} {\bibinfo {author} {\bibfnamefont {C.}~\bibnamefont
  {Nieto}}, \bibinfo {author} {\bibfnamefont {S.~C.}\ \bibnamefont {Blanco}},
  \bibinfo {author} {\bibfnamefont {C.}~\bibnamefont {{Vargas-Garc{\'i}a}}},
  \bibinfo {author} {\bibfnamefont {A.}~\bibnamefont {Singh}}, \ and\ \bibinfo
  {author} {\bibfnamefont {P.~J.}\ \bibnamefont {Manuel}},\ }\bibfield  {title}
  {\enquote {\bibinfo {title} {{{PyEcoLib}}: A python library for simulating
  stochastic cell size dynamics},}\ }\href {\doibase 10.1088/1478-3975/acd897}
  {\bibfield  {journal} {\bibinfo  {journal} {Phys. Biol.}\ }\textbf {\bibinfo
  {volume} {20}},\ \bibinfo {pages} {045006} (\bibinfo {year}
  {2023})}\BibitemShut {NoStop}%
\bibitem [{\citenamefont {Inaba}(2017)}]{inaba2017agestructureda}%
  \BibitemOpen
  \bibfield  {author} {\bibinfo {author} {\bibfnamefont {H.}~\bibnamefont
  {Inaba}},\ }\href {\doibase 10.1007/978-981-10-0188-8} {\emph {\bibinfo
  {title} {Age-{{Structured Population Dynamics}} in {{Demography}} and
  {{Epidemiology}}}}}\ (\bibinfo  {publisher} {Springer},\ \bibinfo {address}
  {Singapore},\ \bibinfo {year} {2017})\BibitemShut {NoStop}%
\bibitem [{\citenamefont {Donnelly}\ and\ \citenamefont
  {Kurtz}(1999)}]{donnelly1999particlea}%
  \BibitemOpen
  \bibfield  {author} {\bibinfo {author} {\bibfnamefont {P.}~\bibnamefont
  {Donnelly}}\ and\ \bibinfo {author} {\bibfnamefont {T.~G.}\ \bibnamefont
  {Kurtz}},\ }\bibfield  {title} {\enquote {\bibinfo {title} {Particle
  {{Representations}} for {{Measure-Valued Population Models}}},}\ }\href
  {\doibase 10.1214/aop/1022677258} {\bibfield  {journal} {\bibinfo  {journal}
  {Ann. Probab.}\ }\textbf {\bibinfo {volume} {27}},\ \bibinfo {pages}
  {166--205} (\bibinfo {year} {1999})}\BibitemShut {NoStop}%
\bibitem [{\citenamefont {Bansaye}\ and\ \citenamefont
  {M{\'e}l{\'e}ard}(2015)}]{bansaye2015stochastic}%
  \BibitemOpen
  \bibfield  {author} {\bibinfo {author} {\bibfnamefont {V.}~\bibnamefont
  {Bansaye}}\ and\ \bibinfo {author} {\bibfnamefont {S.}~\bibnamefont
  {M{\'e}l{\'e}ard}},\ }\href@noop {} {\emph {\bibinfo {title} {Stochastic
  Models for Structured Populations}}},\ Vol.~\bibinfo {volume} {16}\ (\bibinfo
   {publisher} {Springer},\ \bibinfo {year} {2015})\BibitemShut {NoStop}%
\bibitem [{\citenamefont {Matyjaszkiewicz}\ \emph {et~al.}(2017)\citenamefont
  {Matyjaszkiewicz}, \citenamefont {Fiore}, \citenamefont {Annunziata},
  \citenamefont {Grierson}, \citenamefont {Savery}, \citenamefont {Marucci},\
  and\ \citenamefont {{di Bernardo}}}]{matyjaszkiewicz2017bsima}%
  \BibitemOpen
  \bibfield  {author} {\bibinfo {author} {\bibfnamefont {A.}~\bibnamefont
  {Matyjaszkiewicz}}, \bibinfo {author} {\bibfnamefont {G.}~\bibnamefont
  {Fiore}}, \bibinfo {author} {\bibfnamefont {F.}~\bibnamefont {Annunziata}},
  \bibinfo {author} {\bibfnamefont {C.~S.}\ \bibnamefont {Grierson}}, \bibinfo
  {author} {\bibfnamefont {N.~J.}\ \bibnamefont {Savery}}, \bibinfo {author}
  {\bibfnamefont {L.}~\bibnamefont {Marucci}}, \ and\ \bibinfo {author}
  {\bibfnamefont {M.}~\bibnamefont {{di Bernardo}}},\ }\bibfield  {title}
  {\enquote {\bibinfo {title} {{{BSim}} 2.0: {{An Advanced Agent-Based Cell
  Simulator}}},}\ }\href {\doibase 10.1021/acssynbio.7b00121} {\bibfield
  {journal} {\bibinfo  {journal} {ACS Synth. Biol.}\ }\textbf {\bibinfo
  {volume} {6}},\ \bibinfo {pages} {1969--1972} (\bibinfo {year}
  {2017})}\BibitemShut {NoStop}%
\bibitem [{\citenamefont {Dang}, \citenamefont {Grundel},\ and\ \citenamefont
  {Youk}(2020)}]{dang2020cellulara}%
  \BibitemOpen
  \bibfield  {author} {\bibinfo {author} {\bibfnamefont {Y.}~\bibnamefont
  {Dang}}, \bibinfo {author} {\bibfnamefont {D.~A.~J.}\ \bibnamefont
  {Grundel}}, \ and\ \bibinfo {author} {\bibfnamefont {H.}~\bibnamefont
  {Youk}},\ }\bibfield  {title} {\enquote {\bibinfo {title} {Cellular
  {{Dialogues}}: {{Cell-Cell Communication}} through {{Diffusible Molecules
  Yields Dynamic Spatial Patterns}}},}\ }\href {\doibase
  10.1016/j.cels.2019.12.001} {\bibfield  {journal} {\bibinfo  {journal} {Cell
  Systems}\ }\textbf {\bibinfo {volume} {10}},\ \bibinfo {pages} {82--98.e7}
  (\bibinfo {year} {2020})}\BibitemShut {NoStop}%
\bibitem [{\citenamefont {Ma}\ \emph {et~al.}(2022)\citenamefont {Ma},
  \citenamefont {Gowda}, \citenamefont {Anantharaman}, \citenamefont
  {Laughman}, \citenamefont {Shah},\ and\ \citenamefont
  {Rackauckas}}]{ma2022modelingtoolkita}%
  \BibitemOpen
  \bibfield  {author} {\bibinfo {author} {\bibfnamefont {Y.}~\bibnamefont
  {Ma}}, \bibinfo {author} {\bibfnamefont {S.}~\bibnamefont {Gowda}}, \bibinfo
  {author} {\bibfnamefont {R.}~\bibnamefont {Anantharaman}}, \bibinfo {author}
  {\bibfnamefont {C.}~\bibnamefont {Laughman}}, \bibinfo {author}
  {\bibfnamefont {V.}~\bibnamefont {Shah}}, \ and\ \bibinfo {author}
  {\bibfnamefont {C.}~\bibnamefont {Rackauckas}},\ }\href {\doibase
  10.48550/arXiv.2103.05244} {\enquote {\bibinfo {title} {{{ModelingToolkit}}:
  {{A Composable Graph Transformation System For Equation-Based Modeling}}},}\
  } (\bibinfo {year} {2022}),\ \Eprint {http://arxiv.org/abs/2103.05244}
  {arXiv:2103.05244 [cs]} \BibitemShut {NoStop}%
\bibitem [{\citenamefont {Loman}\ \emph {et~al.}(2023)\citenamefont {Loman},
  \citenamefont {Ma}, \citenamefont {Ilin}, \citenamefont {Gowda},
  \citenamefont {Korsbo}, \citenamefont {Yewale}, \citenamefont {Rackauckas},\
  and\ \citenamefont {Isaacson}}]{loman2023catalystc}%
  \BibitemOpen
  \bibfield  {author} {\bibinfo {author} {\bibfnamefont {T.~E.}\ \bibnamefont
  {Loman}}, \bibinfo {author} {\bibfnamefont {Y.}~\bibnamefont {Ma}}, \bibinfo
  {author} {\bibfnamefont {V.}~\bibnamefont {Ilin}}, \bibinfo {author}
  {\bibfnamefont {S.}~\bibnamefont {Gowda}}, \bibinfo {author} {\bibfnamefont
  {N.}~\bibnamefont {Korsbo}}, \bibinfo {author} {\bibfnamefont
  {N.}~\bibnamefont {Yewale}}, \bibinfo {author} {\bibfnamefont
  {C.}~\bibnamefont {Rackauckas}}, \ and\ \bibinfo {author} {\bibfnamefont
  {S.~A.}\ \bibnamefont {Isaacson}},\ }\bibfield  {title} {\enquote {\bibinfo
  {title} {Catalyst: {{Fast}} and flexible modeling of reaction networks},}\
  }\href {\doibase 10.1371/journal.pcbi.1011530} {\bibfield  {journal}
  {\bibinfo  {journal} {PLOS Comput. Biol.}\ }\textbf {\bibinfo {volume}
  {19}},\ \bibinfo {pages} {1--19} (\bibinfo {year} {2023})}\BibitemShut
  {NoStop}%
\bibitem [{\citenamefont {Gillespie}(1977)}]{gillespie1977exacta}%
  \BibitemOpen
  \bibfield  {author} {\bibinfo {author} {\bibfnamefont {D.~T.}\ \bibnamefont
  {Gillespie}},\ }\bibfield  {title} {\enquote {\bibinfo {title} {Exact
  stochastic simulation of coupled chemical reactions},}\ }\href {\doibase
  10.1021/j100540a008} {\bibfield  {journal} {\bibinfo  {journal} {J. Phys.
  Chem.}\ }\textbf {\bibinfo {volume} {81}},\ \bibinfo {pages} {2340--2361}
  (\bibinfo {year} {1977})}\BibitemShut {NoStop}%
\bibitem [{\citenamefont {Lewis}\ and\ \citenamefont
  {Shedler}(1979)}]{lewis1979simulationb}%
  \BibitemOpen
  \bibfield  {author} {\bibinfo {author} {\bibfnamefont {P.~A.~W.}\
  \bibnamefont {Lewis}}\ and\ \bibinfo {author} {\bibfnamefont {G.~S.}\
  \bibnamefont {Shedler}},\ }\bibfield  {title} {\enquote {\bibinfo {title}
  {Simulation of nonhomogeneous poisson processes by thinning},}\ }\href
  {\doibase 10.1002/nav.3800260304} {\bibfield  {journal} {\bibinfo  {journal}
  {Nav. Res. Logist. Q.}\ }\textbf {\bibinfo {volume} {26}},\ \bibinfo {pages}
  {403--413} (\bibinfo {year} {1979})},\ \Eprint
  {http://arxiv.org/abs/https://onlinelibrary.wiley.com/doi/pdf/10.1002/nav.3800260304}
  {https://onlinelibrary.wiley.com/doi/pdf/10.1002/nav.3800260304} \BibitemShut
  {NoStop}%
\bibitem [{\citenamefont {Voliotis}\ \emph {et~al.}(2016)\citenamefont
  {Voliotis}, \citenamefont {Thomas}, \citenamefont {Grima},\ and\
  \citenamefont {Bowsher}}]{voliotis2016stochasticb}%
  \BibitemOpen
  \bibfield  {author} {\bibinfo {author} {\bibfnamefont {M.}~\bibnamefont
  {Voliotis}}, \bibinfo {author} {\bibfnamefont {P.}~\bibnamefont {Thomas}},
  \bibinfo {author} {\bibfnamefont {R.}~\bibnamefont {Grima}}, \ and\ \bibinfo
  {author} {\bibfnamefont {C.~G.}\ \bibnamefont {Bowsher}},\ }\bibfield
  {title} {\enquote {\bibinfo {title} {Stochastic simulation of biomolecular
  networks in dynamic environments},}\ }\href {\doibase
  10.1371/journal.pcbi.1004923} {\bibfield  {journal} {\bibinfo  {journal}
  {PLOS Comput. Biol.}\ }\textbf {\bibinfo {volume} {12}},\ \bibinfo {pages}
  {1--18} (\bibinfo {year} {2016})}\BibitemShut {NoStop}%
\bibitem [{\citenamefont {Thomas}(2018)}]{thomas2018analysisa}%
  \BibitemOpen
  \bibfield  {author} {\bibinfo {author} {\bibfnamefont {P.}~\bibnamefont
  {Thomas}},\ }\bibfield  {title} {\enquote {\bibinfo {title} {Analysis of
  {{Cell Size Homeostasis}} at the {{Single-Cell}} and {{Population Level}}},}\
  }\href {\doibase 10.3389/fphy.2018.00064} {\bibfield  {journal} {\bibinfo
  {journal} {Front. Phys.}\ }\textbf {\bibinfo {volume} {6}} (\bibinfo {year}
  {2018}),\ 10.3389/fphy.2018.00064}\BibitemShut {NoStop}%
\bibitem [{\citenamefont {Thomas}\ and\ \citenamefont
  {Shahrezaei}(2021)}]{thomas2021coordinationc}%
  \BibitemOpen
  \bibfield  {author} {\bibinfo {author} {\bibfnamefont {P.}~\bibnamefont
  {Thomas}}\ and\ \bibinfo {author} {\bibfnamefont {V.}~\bibnamefont
  {Shahrezaei}},\ }\bibfield  {title} {\enquote {\bibinfo {title} {Coordination
  of gene expression noise with cell size: Analytical results for agent-based
  models of growing cell populations},}\ }\href {\doibase
  10.1098/rsif.2021.0274} {\bibfield  {journal} {\bibinfo  {journal} {J. R.
  Soc. Interface}\ }\textbf {\bibinfo {volume} {18}},\ \bibinfo {pages}
  {20210274} (\bibinfo {year} {2021})},\ \Eprint
  {http://arxiv.org/abs/https://royalsocietypublishing.org/doi/pdf/10.1098/rsif.2021.0274}
  {https://royalsocietypublishing.org/doi/pdf/10.1098/rsif.2021.0274}
  \BibitemShut {NoStop}%
\bibitem [{\citenamefont {Parsons}\ \emph {et~al.}(2024)\citenamefont
  {Parsons}, \citenamefont {Bolker}, \citenamefont {Dushoff},\ and\
  \citenamefont {Earn}}]{parsons2024probabilitya}%
  \BibitemOpen
  \bibfield  {author} {\bibinfo {author} {\bibfnamefont {T.~L.}\ \bibnamefont
  {Parsons}}, \bibinfo {author} {\bibfnamefont {B.~M.}\ \bibnamefont {Bolker}},
  \bibinfo {author} {\bibfnamefont {J.}~\bibnamefont {Dushoff}}, \ and\
  \bibinfo {author} {\bibfnamefont {D.~J.~D.}\ \bibnamefont {Earn}},\
  }\bibfield  {title} {\enquote {\bibinfo {title} {The probability of epidemic
  burnout in the stochastic {{SIR}} model with vital dynamics},}\ }\href
  {\doibase 10.1073/pnas.2313708120} {\bibfield  {journal} {\bibinfo  {journal}
  {Proc. Natl. Acad. Sci.}\ }\textbf {\bibinfo {volume} {121}},\ \bibinfo
  {pages} {e2313708120} (\bibinfo {year} {2024})}\BibitemShut {NoStop}%
\bibitem [{\citenamefont {Nardini}\ \emph {et~al.}(2021)\citenamefont
  {Nardini}, \citenamefont {Baker}, \citenamefont {Simpson},\ and\
  \citenamefont {Flores}}]{nardini2021learninga}%
  \BibitemOpen
  \bibfield  {author} {\bibinfo {author} {\bibfnamefont {J.~T.}\ \bibnamefont
  {Nardini}}, \bibinfo {author} {\bibfnamefont {R.~E.}\ \bibnamefont {Baker}},
  \bibinfo {author} {\bibfnamefont {M.~J.}\ \bibnamefont {Simpson}}, \ and\
  \bibinfo {author} {\bibfnamefont {K.~B.}\ \bibnamefont {Flores}},\ }\bibfield
   {title} {\enquote {\bibinfo {title} {Learning differential equation models
  from stochastic agent-based model simulations},}\ }\href {\doibase
  10.1098/rsif.2020.0987} {\bibfield  {journal} {\bibinfo  {journal} {J. R.
  Soc. Interface}\ }\textbf {\bibinfo {volume} {18}},\ \bibinfo {pages}
  {20200987} (\bibinfo {year} {2021})}\BibitemShut {NoStop}%
\bibitem [{\citenamefont {Tankhilevich}\ \emph {et~al.}(2020)\citenamefont
  {Tankhilevich}, \citenamefont {{Ish-Horowicz}}, \citenamefont {Hameed},
  \citenamefont {Roesch}, \citenamefont {Kleijn}, \citenamefont {Stumpf},\ and\
  \citenamefont {He}}]{tankhilevich2020gpabc}%
  \BibitemOpen
  \bibfield  {author} {\bibinfo {author} {\bibfnamefont {E.}~\bibnamefont
  {Tankhilevich}}, \bibinfo {author} {\bibfnamefont {J.}~\bibnamefont
  {{Ish-Horowicz}}}, \bibinfo {author} {\bibfnamefont {T.}~\bibnamefont
  {Hameed}}, \bibinfo {author} {\bibfnamefont {E.}~\bibnamefont {Roesch}},
  \bibinfo {author} {\bibfnamefont {I.}~\bibnamefont {Kleijn}}, \bibinfo
  {author} {\bibfnamefont {M.~P.~H.}\ \bibnamefont {Stumpf}}, \ and\ \bibinfo
  {author} {\bibfnamefont {F.}~\bibnamefont {He}},\ }\bibfield  {title}
  {\enquote {\bibinfo {title} {{{GpABC}}: A {{Julia}} package for approximate
  {{Bayesian}} computation with {{Gaussian}} process emulation},}\ }\href
  {\doibase 10.1093/bioinformatics/btaa078} {\bibfield  {journal} {\bibinfo
  {journal} {Bioinformatics}\ }\textbf {\bibinfo {volume} {36}},\ \bibinfo
  {pages} {3286--3287} (\bibinfo {year} {2020})}\BibitemShut {NoStop}%
\bibitem [{\citenamefont {Danisch}\ and\ \citenamefont
  {Krumbiegel}(2021)}]{danisch2021makiea}%
  \BibitemOpen
  \bibfield  {author} {\bibinfo {author} {\bibfnamefont {S.}~\bibnamefont
  {Danisch}}\ and\ \bibinfo {author} {\bibfnamefont {J.}~\bibnamefont
  {Krumbiegel}},\ }\bibfield  {title} {\enquote {\bibinfo {title} {Makie.jl:
  {{Flexible}} high-performance data visualization for {{Julia}}},}\ }\href
  {\doibase 10.21105/joss.03349} {\bibfield  {journal} {\bibinfo  {journal} {J.
  Open Source Softw.}\ }\textbf {\bibinfo {volume} {6}},\ \bibinfo {pages}
  {3349} (\bibinfo {year} {2021})}\BibitemShut {NoStop}%
\end{thebibliography}%

\end{document}